\documentclass[aip,pop,preprint,a4paper]{revtex4-1}

\usepackage{amsmath}
\usepackage{amssymb}
\usepackage{graphicx}
\usepackage{dcolumn}
\usepackage{bm}

\draft 

\begin{document}


\title{Gyrosymmetry: global considerations} 



\author{J. W. Burby}
 \affiliation{Princeton Plasma Physics Laboratory, Princeton, New Jersey 08543, USA}
\author{H. Qin}
 \affiliation{Princeton Plasma Physics Laboratory, Princeton, New Jersey 08543, USA}
 \affiliation{Dept. of Modern Physics, University of Science and Technology of China, Hefei, Anhui 230026, China}


\date{\today}

\begin{abstract}
In the guiding center theory, smooth unit vectors perpendicular to the magnetic field are required to define the gyrophase. The question of global existence of these vectors is addressed using a general result from the theory of characteristic classes. It is found that there is, in certain cases, an obstruction to global existence. In these cases, the gyrophase cannot be defined globally. The implications of this fact on the basic structure of the guiding center theory are discussed. In particular it is demonstrated that the guiding center asymptotic expansion of the equations of motion can still be performed in a globally consistent manner when a single global convention for measuring gyrophase is unavailable. The latter fact is demonstrated directly by deriving a new expression for the guiding-center Poincar\'e-Cartan form exhibiting no dependence on the choice of perpendicular unit vectors.
\end{abstract}


\maketitle 

\section{Introduction}
There is no doubt that the Hamiltonian formulation of guiding center theory is a foundational aspect of modern gyrokinetic theories. Simply put, it provides a means for deforming the single-particle phase space so as to illuminate the approximate symmetry associated to the magnetic moment, the gyrosymmetry, while keeping the Hamiltonian structure of the particle dynamics in focus. However, in spite of its importance and the number of years it has been studied \cite{kruskal58,northrop,littlejohn81,littlejohn83,madsen10,WB,brizard95,banos,kauf92,carybrizard,hahmbrizard,krommes83}, there are still poorly understood subtleties in the theory.

In this paper, we study the subtleties associated with the so-called ``perpendicular unit vectors'' that make an appearance in virtually every version of the theory \cite{KBM,kruskal58,northrop,littlejohn81,littlejohn83,madsen10,WB,brizard95,banos,kauf92,carybrizard,hahmbrizard}. These quantities, hereafter referred to as $e_1$ and $e_2$, are smooth unit vector fields everywhere perpendicular to the magnetic field and to one another, meaning they form an orthonormal triad together with $b=B/||B||$ in the velocity space. From one point of view, they appear in the formalism for the sake of identifying an angular variable $\theta$, the gyrophase, that evolves on a fast timescale with respect to the evolution timescale of the remaining dynamical variables, thereby putting the guiding center problem in the setting of the generalized method of averaging described in Ref. \onlinecite{KBM}. In particular, when the equations of motion for a strongly magnetized charged particle are expressed using a cylindrical parameterization of velocity space such that the cylindrical axis points along the magnetic field, then it can be shown that the polar angle associated to this cylindrical coordinate system furnishes such a fast angle. This angle is measured with respect to a pair of mutually orthogonal normalized vectors $e_1,e_2$ lying in the plane perpendicular to $B$. Because the magnetic field varies spatially, $e_1,e_2$ must also vary in space so as to accommodate the constraint $e_1\cdot B=0$.
\begin{figure}
\includegraphics{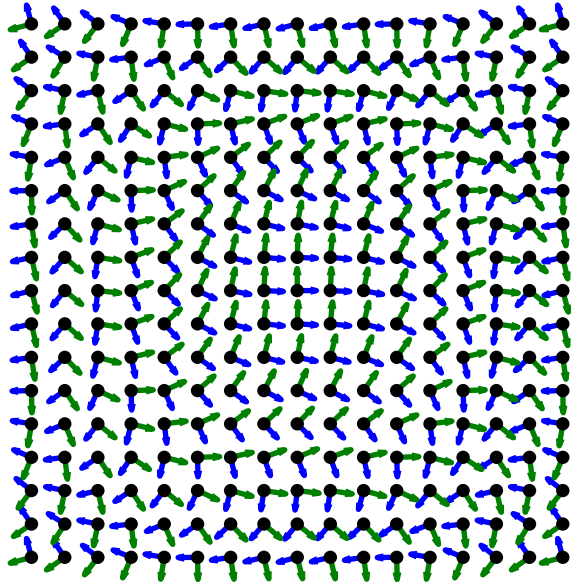}
\caption{\label{perp}A typical arrangement of the perpendicular unit vectors $e_1,e_2$ for a uniform magnetic field that points out of the page. The two sets of arrows represent $e_1$ and $e_2$. While in this case, $e_1$ and $e_2$ are not \emph{required} to vary in space, for a more general sort of magnetic field, they would be. Reprinted from Phys. Plasmas 19, 052106 (2012). Copyright 2012 American Institute of Physics.}
\end{figure}
 Therefore these $e_1,e_2$ furnish an example of perpendicular unit vectors (see Fig. \ref{perp}). From another, more geometric point of view, the perpendicular unit vectors usher themselves into the formalism so as to facilitate parameterizing the zero'th-order symmetry loops, or Kruskal Rings \cite{kruskal62,qinRINGS,qin07} associated with the gyrosymmetry; one of the vectors, say $e_1$, distinguishes a point on each Kruskal Ring which then serves as a reference or zero angle. Interestingly, nobody's version of the theory ever provides a general, constructive definition of these $e_1,e_2$ in terms of known quantities. This is the first hint that there is more to these vector fields than meets the eye.

Perhaps the reason nobody provides such a definition is that, in the most general setting where the guiding center expansion applies, $e_1,e_2$ simply cannot be defined globally, that is, there might not even be one vector field defined over the \emph{entire} configuration space that is at once perpendicular to $B$ and of unit length. While it is easy to see that smooth $e_1,e_2$ can always be defined \emph{locally} in some, generally tiny, open neighborhood of any point $p$ in the configuration space\footnote{One of the members of the standard basis for $\mathbb{R}^3$ must be non-parallel to $B(p)$. Let that member be denoted $e$. Now consider the function $g(x)=B(x)\times e$. $g$ is obviously smooth, non-zero when $x=p$, and perpendicular to $B$ whenever it is not zero. The continuity of $g$ guarantees that if $g$'s domain is restricted to a small enough open neighborhood $U$ of $p$, then it will be nonzero at every point in that domain. Thus, on $U$ the formula $e_1(u)=g(u)/|g(u)|$ defines a smooth unit vector perpendicular to $B$.}, this in no way implies that these locally defined perpendicular unit vectors extend to well-defined global quantities \cite{sug,comment,response}. So could there be an obstruction the global existence of smooth $e_1,e_2$ in some cases?


If we take this question seriously, a more important one arises immediately. Is the guiding center theory still valid without global perpendicular unit vectors? As the theory is carried out to higher order, expressions involving the perpendicular unit vectors and their derivatives appear in the equations of motion; see Ref. \onlinecite{littlejohn81} for instance. So it might seem plausible that the existence of global equations of motion is tied to the global properties of $e_1,e_2$. 

Here we will put both of these questions to rest. We will provide a complete mathematical description of the obstruction to global perpendicular unit vectors and show that this obstruction does not always vanish. However, we demonstrate that the obstruction does indeed vanish if the physical domain is an open solid torus. Then we will show that the guiding center theory \emph{does} provide consistent global equations of motion in the absence of global $e_1,e_2$ owing to the fact that the symmetry associated with the magnetic moment  is always globally defined. To illustrate this second point, we provide an expression for the guiding center Poincar\'e-Cartan one-form in terms of globally defined physical quantities like $B$; neither the perpendicular unit vectors nor the gyrophase appear in the expression.  

The paper is structured as follows. In \ref{one} we provide a simple example of a magnetic field that does not admit global $e_1,e_2$. Then in \ref{two}, we provide a complete mathematical description of the obstruction to global perpendicular unit vectors in the most general case. As an example illustrating the theory, we prove in \ref{three} that if the physical domain\footnote{In the context of the guiding center theory, at least the regions where $b$ cannot be defined must be excluded from the physical domain.}, $D$,  particles are tracked through is an open, solid torus, then it is always possible to find global $e_1,e_2$. This is even true when the magnetic field lines are chaotic! Then we give a non-trivial example of a magnetic field that does not admit global $e_1,e_2$.  Finally in \ref{four}, we show that the guiding center theory \emph{does} provide consistent global equations of motion in the absence of global $e_1,e_2$.

\section{\label{one}A motivating example: the field due to a magnetic monopole}
The field due to a magnetic monopole provides probably the simplest illustration of the obstruction to the existence of global $e_1,e_2$. Perhaps the simplicity comes at the cost of physical relevance, but the latter will be reclaimed later after developing some machinery. Amusingly, the possibility that this example \emph{is} physically relevant has never been ruled out. See Ref. \onlinecite{polchinski} for an interesting discussion of the current status of magnetic monopoles in theoretical physics.

\begin{figure}
\includegraphics{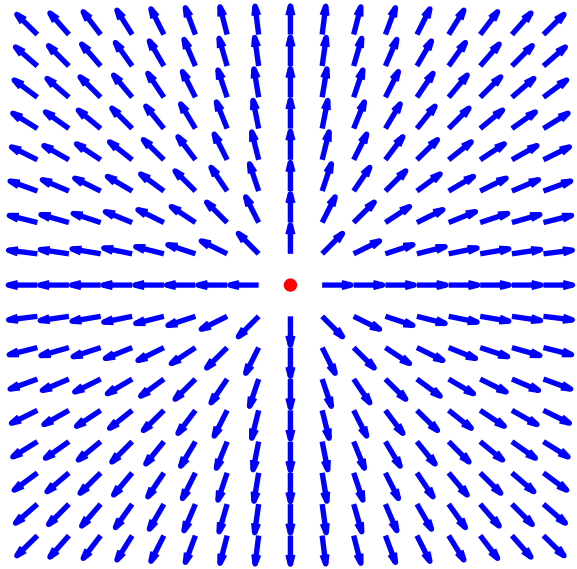}
\caption{\label{monopole}The magnetic field due to a magnetic monopole. Note that $\nabla\cdot B=0$ except at the origin, which is depicted as a large central dot. Reprinted from Phys. Plasmas 19, 052106 (2012). Copyright 2012 American Institute of Physics.}
\end{figure}

The monopole field is given by 
\begin{align}
B(x)=\frac{1}{||x||^2}e_r(x),
\end{align}
where $e_r$ is the radial unit vector from a spherical coordinate system about the origin. It is depicted in Fig. \ref{monopole}.
Sufficiently far from the singularity at the origin, we could in principle develop the guiding center approximation. So let the physical domain $D$ where particles would move consist of the open region exterior to some sphere of radius $r_o$ centered on the origin. Now we will check if there is a perpendicular unit vector defined on all  of $D$.

If there were such a vector field, $e_1$, then it could be restricted to a sphere centered on the origin with radius $r_a>r_o$, $S_{r_a}$. Because $B|S_{r_a}$ is parallel to the vector normal to $S_{r_a}$, $e_1|S_{r_a}$ would have to be everywhere tangent to $S_{r_a}$. Thus, 
\[
e_1|S_{r_a}:S_{r_a}\rightarrow TS_{r_a},
\] 
where $ TS_{r_a}$ denotes the tangent bundle \cite{FoM} of $S_{r_a}$, would furnish an example of a smooth non-vanishing tangent vector field on the sphere. But this situation is impossible by the famous ``hairy ball theorem''. It follows that no such $e_1$ exists.

There are two essential features of this example. First of all, notice that $D$ has a ``hole'' due to excluding the region with $r<r_o$, thus giving $D$ the shape of a peach without the pit. If instead $D$ were chosen to be some solid spherical region separated from the singularity at the origin, then it \emph{would} be possible to find an $e_1$ (we won't prove this now). But then $D$ would be hole free. So we see that the obstruction to the existence of $e_1$ is somehow related to the topology of $D$, in particular the presence of holes (or lack thereof) is important. Second, notice that the utility of the hairy ball theorem derives entirely from the fact that the planes perpendicular to the magnetic field are arranged to be tangent to the spheres $S_{r_a}$. Thus this distribution of perpendicular planes impacts the existence of global perpendicular unit vectors. In particular, note that even if $D$ had holes, were the monopole field replaced with a uniform field, then global perpendicular unit vectors \emph{could} be found.

\section{\label{two}The general obstruction to global perpendicular unit vectors} 
Now consider the general problem of finding a perpendicular unit vector defined on the entirety of an arbitrary physical domain $D$. $D$ can have any number of holes, but we will insist that it be an open subset of $\mathbb{R}^3$ and that the magnetic field in this region is never zero. Thus $D$ might have the appearance of a block (not just a slice) of swiss cheese. In practice, $D$ would be determined by first choosing a domain where particles will move, and then removing those regions where the guiding-center ordering breaks down.

We claim that the key ingredients in the solution to this existence problem are the hole structure of $D$ and the divergence-free vector field $N$ discussed in depth by Littlejohn \cite{littlejohn81,littlejohn83},
\begin{align}\label{N}
N=&\frac{1}{2}b\bigg(\text{Tr}(\nabla b\cdot \nabla b)-(\nabla\cdot b)^2\bigg)\\
&+(\nabla\cdot b)b\cdot\nabla b-b\cdot\nabla b\cdot \nabla b.\nonumber
\end{align}
In particular, in order for a global perpendicular unit vector to exist, it is both \emph{necessary and sufficient} that there be zero net flux of $N$ through each boundary-free surface encapsulating a hole in $D$.

Using just Stoke's theorem, it is easy to see that the latter condition is indeed \emph{necessary} for global existence. If there were a globally defined $e_1$, then the vector $R=(\nabla e_1)\cdot (b\times e_1)$ would be globally defined. It is straightforward to show\cite{littlejohn81,littlejohn83} that this implies $N$ has a globally defined vector potential $N=\nabla\times R$. Therefore, if $S$ were a boundary-free surface encapsulating a hole in $D$,
\begin{align}
\int_SN\cdot dA=\int_S\nabla\times R\cdot dA=\int_{\partial S}R\cdot dl=0.
\end{align} 

To show \emph{sufficiency} is not nearly as simple. Unfortunately, a properly rigorous demonstration would require a lengthy digression into the theory of principal bundles and characteristic classes, topics that are discussed by a master of these subjects in Ref. \onlinecite{chern}. While we will make no attempt to provide the complete digression, we have included an appendix describing how the theorem on page 118 of the last reference can be applied to our existence problem to prove sufficiency of our flux condition. More curious readers will want to read Ref. \onlinecite{chern} in detail.

Regardless of how the flux condition is proved, however, it is helpful to understand the following physical argument for why it is \emph{feasible}. As already discussed by Littlejohn in Ref. \onlinecite{GGI}, $N$ can be interpreted as a kind of magnetic field whose coupling constant is the magnetic moment (instead of the electric charge). If the flux condition on $N$ is not satisfied, then because $\nabla\cdot N=0$ on $D$, then there must be \emph{monopole sources} for the field $N$ lurking in $D$'s holes,
\begin{align}
\int_SN\cdot dA=2\pi Q_{\text{gyro}}.
\end{align}
Here $Q_{\text{gyro}}$ we term the \emph{gyrokinetic monopole charge} contained in the hole encapsulated by $S$. A striking fact, which pushes the $N$-magnetic field analogy even further, is that $Q_{\text{gyro}}$ must be an integer. The latter can of course be identified with Dirac's quantization condition on the magnetic charge, a point also discussed in Ref. \onlinecite{chern}. Now recall that when tracking the evolution of the quantum phase of an electron outside of a Dirac monopole, a single global convention for measuring this phase is impossible; there must be at least two distinct measurement conventions, corresponding to the domains where the vector potential can be defined without singularities. In gyrokinetics, the gyrophase can be considered an analogue of the quantum phase and $R$ an analogue of the vector potential. To make this analogy precise again requires discussing principal bundles. However, because gyrophase and quantum phase represent redundant physical information, albeit in different contexts, it is perhaps reasonable on physical grounds. Thus it should not be surprising that a single convention for measuring gyrophase, corresponding to a choice of perpendicular unit vector, is unavailable when $D$ encapsulates gyrokinetic monopoles. Likewise, because there are no issues defining the quantum phase when an electron's physical domain does not encapsulate Dirac monopoles, it should not be surprising that there are not issues defining the gyrophase when $D$ does not encapsulate gyrokinetic monopoles.

\section{\label{three}Some example assessments of the existence of global perpendicular unit vectors}
Now we will apply the machinery developed in the previous section to assess the existence of global perpendicular unit vectors for a few example choices of $D$ and $B$. Because of their relevance to magnetic confinement, we will first treat the broad class of examples where $D$ is an open solid torus and $B$ is only constrained to be non-vanishing on $D$. We will show that, in these examples, global perpendicular unit vectors can always be found. Then we will consider a more exotic example where $B$ is linear and vanishes at a single point and $D$ is taken to be the region surrounding the field null. In this case global perpendicular unit vectors do \emph{not} exist.

When $D$ is an open solid torus, for instance the region contained within the vacuum vessel of a tokamak device, then it is intuitively clear that every boundary-less 2-dimensional surface contained in $D$ arises as the boundary of some 3-dimensional region. This statement can of course be demonstrated rigorously using some basic results from algebraic topology\cite{bott}. Stoke's theorem then implies that, because $\nabla \cdot N=0$, the flux of $N$ through any such surface must vanish. Therefore we arrive at the following conclusion: when $D$ is an open solid torus, global perpendicular unit vectors always exist.

It is worth mentioning that this conclusion holds even when there are chaotic magnetic field lines. To see that this is \emph{reasonable}, consider a typical tokamak field that has been subjected to a resonant magnetic perturbation. Often, for instance in Ref. \onlinecite{RMP}, these perturbations are not large enough to completely kill the toroidal component of the magnetic field at any point within the last closed flux surface (assume this region is $D$). However, it is will known that they may nonetheless create regions of chaotic field lines. Therefore, in spite of the presence of chaotic field lines, the vector
\begin{align}
E_1=e_R\times B=B_\phi e_z-B_z e_\phi, 
\end{align}
where $e_R,e_\phi$ are the cylindrical radial and azimuthal unit vectors, vanishes nowhere in $D$ and so defines a global perpendicular unit vector $e_1=E_1/||E_1||$. Similarly ``X-points'' and ``O-points'' lead to no obstruction to a global $e_1,e_2$.

Now consider the magnetic field given by 
\begin{align}
B(x,y,z)=y e_x+ze_y+xe_z.
\end{align}
Let $D=\mathbb{R}^3 \setminus S_{r_o}$, where $S_{r_o}$ is a solid sphere centered on the origin whose radius is much larger than any gyroradius of interest. Thus we exclude from $D$ the only region where the gyrocenter coordinate system cannot be treated perturbatively. Note that there is nothing singular about $B$ at $0$ even though $b$ is. Also note that the current density $\nabla\times B$ is uniform. It is straightforward to compute the flux of $N$ through a sphere of any radius centered on the origin, which turns out to be $-4\pi$. This implies there is a gyrokinetic monopole charge $Q_{\text{gyro}}=-2$ contained in $D$'s hole. This rules out the possibility of the existence of a perpendicular unit vector defined on all of $D$. Note that we could not have proven this last result by appealing directly to the hairy ball theorem; instead we had to utilize the more general flux condition.

\section{\label{four}How the guiding center theory works without global perpendicular unit vectors}
When a perpendicular unit vector cannot be defined globally, the usual notion of gyrophase looses its global meaning as well. So what happens to the guiding center perturbation expansion? Because $D$ can always be covered by (perhaps tiny) open regions $U_\alpha$ in which \emph{local} $e_1,e_2$ are defined, the perturbation procedure can certainly be carried out in each of these patches. The result of each of these local calculations would then consist of formal phase space coordinate changes given as formal one-to-one maps $\phi_\alpha:U_\alpha\times\mathbb{R}^3\rightarrow \mathbb{R}^6$ that lead to simpler equations of motion in the new coordinates. However, these coordinate changes will not necessarily fit together to define a global coordinate change, i.e. an invertible mapping of the entire phase space into itself. Therefore, when calculating the motion of a particle as it moves from one $U_\alpha$ to the next, it becomes necessary to occasionally pass the mechanical state from one $\phi_\alpha$ to another in order to continue using the simplified equations of motion provided by the perturbation theory. While this can be done formally by developing expressions for $\phi_\alpha\circ\phi_\beta^{-1}$, practically it would involve truncating asymptotic series each time the particle crossed from one $U_\alpha$ to the next. This could lead to coherently accumulating error in a simulation, and, in general, would destroy the Hamiltonian properties of the simplified equations of motion.

A far better approach is to look for a global change of coordinates to accomplish the perturbation theory \emph{from the outset}. This way the difficulties associated with truncating the expansions of the $\phi_\alpha\circ\phi_\beta^{-1}$ could be avoided altogether. We have found that such a global coordinate change \emph{can} be found for the guiding center problem owing essentially to the fact that the zero'th order symmetry is globally defined. We arrived at this conclusion by applying a version of Lie perturbation theory to the guiding center problem that synthesizes Littlejohn's Poincar\'e-Cartan one-form approach developed in Ref. \onlinecite{littlejohn82} (also see Ref. \onlinecite{cary81}) with the group-theoretic structure provided by a zero'th order symmetry. Littlejohn's formalism provided the means for performing the perturbation expansion in each of the regions of phase space where the perpendicular unit vectors can be defined, while the globally defined symmetry served as the needle that sews these local calculations into a global result. 

Because the mathematical formalism we used to arrive at this conclusion draws heavily on fiber bundle theory, we will not reproduce our method of proof here. The key point, however, is simple. Because the coordinate change used in the perturbation theory is defined in terms of the flow map of a Lie generator, i.e. a vector field, the coordinate change will be globally defined if and only if the Lie generator is globally defined. A Lie generator will be globally defined if and only if its local expressions transform as a vector should upon changing from one local coordinate system to another. If the coordinate systems we use on phase space consist of locally defined cylindrical velocity space parameterizations, corresponding to different local conventions for measuring the gyrophase, then \emph{the vector transformation law simplifies to the condition for gyrogauge invariant Lie generators}\cite{GGI}. Thus, provided gyrogauge invariant Lie generators are used, the coordinate change derived in the perturbation theory will be globally defined as desired. Readers interested in a more detailed discussion can contact one of us via email.

This fact has the happy consequence that, provided gyrogauge invariant Lie generators are employed, the guiding- or gyro- center Poincar\'e-Cartan one-form must be a globally defined quantity even when the perpendicular unit vectors are not.Therefore, if we work in a globally defined coordinate system, such as the obvious cartesian position and velocity coordinates, the Poincar\'e-Cartan one-form will be manifestly independent of the perpendicular unit vectors. We will demonstrate this explicitly to drive home the point that the guiding center theory will work even without global perpendicular unit vectors. 

For simplicity we will only consider the time-independent case. Let $\mathbf{A}$ denote the magnetic vector potential vector field and $\mathbf{B}=\nabla\times \mathbf{A}$ denote the magnetic field. Then the Poincar\'e-Cartan one-form, ordered in one of the standard ways \cite{northrop}, is given by
\begin{align}
\vartheta_\epsilon(x,v,t)&=\mathbf{A}(x)\cdot dx+\epsilon v\cdot dx -\epsilon^2 \frac{1}{2}v\cdot v dt\\
&=\vartheta^0+\epsilon\vartheta^1+\epsilon^2\vartheta^2.\nonumber
\end{align}
One can consider all variables dimensionless or not. In the latter case, $\mathbf{A}$ should be considered to be normalized by the charge-to-mass ratio of the particle in question so that $\nabla\times \mathbf{A}$ has the units of frequency. The coordinates used in this expression are cartesian position and velocity, $(x,v)$.

This one-form defines the dynamical vector field $X_\epsilon(x,v,t)$ through the formula
\begin{align}
X_\epsilon\lrcorner d\vartheta_\epsilon=0.
\end{align}
It is straightforward to verify that this implies
\begin{align}
\dot x(x,v)&=\epsilon v\\
\dot v(x,v)&=v\times \mathbf{B}(x).
\end{align}
Now, applying Littlejohn's gyrogauge invariant Poincar\'e-Cartan perturbation theory in a domain of phase space where we have a locally defined perpendicular unit vector as in Ref. \onlinecite{GGI}, the truncated Poincar\'e-Cartan one-form becomes
\begin{align}\label{lj}
\hat\vartheta(x,v_\parallel,v_\perp,t&)=\bigg(\mathbf{A}(x)+v_\parallel b(x)\bigg)\cdot dx\\
+\frac{1}{2}\frac{v_\perp^2}{||\mathbf{B}(x)||}&\bigg(d\theta-R(x)\cdot dx\bigg)-\bigg(\frac{1}{2}v_\parallel^2+\frac{1}{2}v_\perp^2\bigg)dt,\nonumber
\end{align}
which involves the unphysical $e_1,e_2$ through $R=(\nabla e_1)\cdot e_2$. The coordinates used in this expression are cartesian position $x$ and cylindrical velocity coordinates $(v_\perp,v_\parallel,\theta)$, where $\theta$ is measured with respect to the local perpendicular unit vector $e_1$. 

Now we simply change back to cartesian position and velocity coordinates according to the mapping
\begin{align*}
v=&v_\parallel b(x)+v_\perp\cos(\theta) e_1(x)-v_\perp\sin(\theta) b(x)\times e_1(x)\\
x=&x.
\end{align*}
Without displaying the calculation, the one-form then takes the form
\begin{align}\label{glo}
\hat\vartheta(x,v,t)=&\bigg(\mathbf{A}(x)+v\cdot b(x) b(x)\bigg)\cdot dx \\
+\frac{1}{2}\frac{(\Pi(x)\cdot v)^2}{||\mathbf{B}(x)||}&\bigg[\bigg(\nabla b\cdot\frac{b(x)\times vv\cdot b(x)}{||b(x)\times v||^2}\bigg)\cdot dx\nonumber\\
&~~-\bigg(\frac{b(x)\times v}{||b(x)\times v||^2}\bigg)\cdot dv\bigg]-\frac{1}{2}v\cdot v dt\nonumber,
\end{align}
where $\Pi(x)=1-b(x)b(x)$ is the perpendicular projection tensor. Clearly the perpendicular unit vectors appear nowhere in the expression. Furthermore, it has exactly the same symmetry properties as Eq.\,(\ref{lj}) because no approximations were made passing from that expression to this one. In particular, the dynamical equations implied by the new expression must conserve the magnetic moment exactly. 

Boghosian\cite{bog} has achieved a similar result previously in the relativistic context. However, he decided to introduce extra variables with compensatory Lagrange multipliers presumably in order to continue to work with the parallel and perpendicular velocity as coordinates. Thus the above expression is indeed a distinct and, to our knowledge, new result. One point regarding its derivation is especially important: if $R$ were to be neglected in Eq.\,(\ref{lj}), the dependence of the one-form on the gyrophase convention would \emph{not} disappear upon passing to cartesian position and velocity coordinates. We would also like to mention that we have recently been informed \cite{zhi} of a Lie perturbation method that succeeds in attaining \ref{glo} directly, without ever resorting to the cylindrical velocity space parameterization.

\section{Conclusion and discussion}
To summarize, we have identified the obstruction to the global existence of perpendicular unit vectors in terms of a flux condition on the nameless vector $N$. Through examples, we showed that this obstruction does not trivially vanish in all cases. In particular, we have given two simple examples where the guiding center ordering is applicable, but global perpendicular unit vectors fail to exist. However, we demonstrated that when the physical domain particles move through is an open solid torus, global perpendicular unit vectors always exist. We have also provided a physically plausible explanation for the flux condition in terms the new concept of gyrokinetic monopoles. 

Then we proceeded to explain how the guiding center theory works when global perpendicular unit vectors are unavailable. In particular, we derived an expression for the guiding-center Poincar\'e-Cartan one-form in a coordinate system rectilinear in both position and velocity only involving globally defined quantities.

Looking at what we have done from a practical point of view, we have identified some difficulties researchers will face when trying to simulate gyrophase-dependent dynamics\cite{gyrogauge,yu09,kolesnikov} in configurations where global perpendicular unit vectors cannot be defined. When dealing with such deviant cases numerically, for instance in a particle-in-cell simulation, it will be necessary to either define a number of gyrophase conventions that cover the phase space and keep track of which of these ``patches'' particles live in, or resort to the global expression for the Poincar\'e-Cartan one-form given at the end of the previous section. In the former case, care must be taken to avoid spending too much time keeping track of a particle's ``patch'', while in the latter case this could be avoided. However, the cost incurred by using the global version of the one-form comes in the form of complicated equations of motion. While simulations of the interior of tokamaks should be able to avoid multiple gyrophase conventions by finding global perpendicular unit vectors (which must exist), this may not be the case in configurations such as the polywell\cite{carr} that involve field nulls in the region of interest. Around each of these field nulls, bubble-like regions must be excluded from the physical domain to ensure the validity of treating the gyrocenter coordinate system perturbatively. If gyrokinetic monopole charge resides in any of these cavities, then perpendicular unit vectors will be unavailable in the ``safe'' region exterior to these cavities.

%
%

%

\begin{acknowledgments}
This work was supported by the U.S. Department of Energy under contract number DE-AC02-09CH11466.
\end{acknowledgments}

\appendix
\section{Principal circle bundles}
Here we define and discuss the notion of principal circle bundle. A more complete exposition can be found in Ref. \onlinecite{sharpe}. First some terminology.  Let $P$ be a manifold and $\Phi:S^1\times P\rightarrow P$ a smooth map, where $S^1=\mathbb{R}~\text{mod}~2\pi$ denotes the circle. If $\theta_1,\theta_2\in S^1$, then we take the symbol $\theta_1+\theta_2$ to mean addition modulo $2\pi$. For a fixed $\theta\in S^1$ define the map $\Phi_\theta:P\rightarrow P$ by the formula $\Phi_\theta(p)=\Phi(\theta,p)$, where $p$ is any point in $P$. $\Phi$ is said to be a left circle action when $\Phi_{\theta_1}\circ\Phi_{\theta_2}=\Phi_{\theta_1+\theta_2}$ and $\Phi_0$ is the identity on $P$. Given a point $p\in P$, the set $\mathcal{O}_p=\{p^\prime\in P|\exists\theta\in S^1~\text{s.t.}~\Phi_\theta(p)=p^\prime\}$ is called the orbit of $\Phi$ through $p$. A left circle action is said to be free if $\Phi_\theta(p)=p$ if and only if $\theta=0$. Intuitively, a left circle action is free if when the second argument of $\Phi$ is held fixed at $p_o$, the resulting map establishes a one-to-one correspondence between the orbit through $p_o$ and the circle. A \emph{principal circle bundle} is a manifold $P$ together with a free left circle action $\Phi:S^1\times P\rightarrow P$. If there is a manifold $B$ and a smooth map $\pi:P\rightarrow B$ such that $\pi$ is surjective, its Jacobian matrix has full rank at each point $p\in P$, and $\pi^{-1}(b)$ is an entire orbit for each $b\in B$, then $P/S^1\equiv B$ is referred to as the base of the principal circle bundle $P$ and $\pi$ is referred to as the bundle projection map. Because it can be shown \cite{FoM} such a $B$ and $\pi$ can always be found for a principal circle bundle, the following intuitive picture of such bundles emerges. A principal circle bundle is nothing more than a collection of circles (the orbits) smoothly parameterized by the base $P/S^1$.

There is a subtle aspect of this picture however. Notice that while it is possible to fix a point $p_o\in P$ as the second argument in $\Phi$ and establish a correspondence between the orbit through $p_o$ and $S^1$, if $\Phi_\theta(p_o)$ were used in place of $p_o$, the result would be a \emph{different} correspondence between the same two objects $\mathcal{O}_{p_o}$ and $S^1$. This is because $\mathcal{O}_{p_o}=\mathcal{O}_{\Phi_\theta(p_o)}$. Therefore, while the orbits $\mathcal{O}_p$ ``look'' like distorted copies of the circle, they lack a natural choice for the $0$, or reference angle. 

On the other hand, it is often convenient take a bunch of nearby orbits and smoothly assign to each of them a reference point so that each point on this bunch of orbits can be assigned an angle in an unambiguous way. Such an assignment of reference points is called a local section. Formally, given an open subset $U_\alpha\subset P/S^1$ of the base, a local section $s_\alpha:U_\alpha\rightarrow\pi^{-1}(U_\alpha)$ is a mapping from $U_\alpha$ into the collection of orbits that project onto $U_\alpha$ that satisfies the equation $\pi\circ s_\alpha=\text{id}_{U_\alpha}$, which simply says that $s_\alpha$ assigns a single point to each of the orbits ``attached'' to $U_\alpha$. Local sections can always be found. However, a global section $s:P/S^1\rightarrow P$, which would smoothly assign a reference point to \emph{all} of the orbits that make up $P$, may not exist. If a global section does exist, then the principal bundle is referred to as being \emph{trivial}.

In the presence of a local section, the process of assigning an angle to each point in the bunch of orbits attached to $U_\alpha$ can be formalized as a special coordinate system on $\pi^{-1}(U_\alpha)$ known as a bundle chart. If $p\in \pi^{-1}(U_\alpha)$, then, because the action is free, there is a unique $g_\alpha(p)\in S^1$ such that
$
p=\Phi_{g_\alpha(p)}s_\alpha(\pi(p)).
$ 
This defines the functions $g_\alpha:\pi^{-1}(U_\alpha)\rightarrow S^1$. The bundle charts $\phi_\alpha:\pi^{-1}(U_\alpha)\rightarrow U_\alpha\times S^1$ are then given by the formula
$
\phi_\alpha(p)=(\pi(p),g_\alpha(p)).
$ By this definition, when looking at a principal circle bundle locally in a bundle chart, it looks like a bunch of bike tires hanging on a multi-dimensional horizontal rod. The orbits are the tires while the base is the rod. It is also useful to think of the bundle charts as ``symmetry-aligned'' coordinate systems, where the symmetry is defined by $\Phi$.

\section{Principal connections}

This appendix gives the definition of a principal connection and briefly explores some of the basic properties of these objects relevant to this article. A much more thorough discussion can be found in Ref. \onlinecite{naber}. 

Given a principal circle bundle $(P,\Phi)$ and a real number $\xi$, the \emph{infinitesimal generator} $\xi_P$ associated to $\xi$ is the vector field on $P$ given by $\xi_P(p)=\frac{d}{d\theta}\big|_{\theta=0}\Phi_{\xi\theta}(p)$. So $\xi_P$ points in the direction of the symmetry associated with $\Phi$. A \emph{principal connection}, or connection form on $P$ is a one-form, $\mathcal{A}$, with the following two properties:
\begin{align*}
1)& ~\forall \xi\in\mathbb{R},~~\mathcal{A}(\xi_P)=\xi\\
2)& ~\forall\theta\in S^1,~~\Phi_{\theta}^*\mathcal{A}=\mathcal{A}.
\end{align*}

Connection forms have a useful local structure when viewed in the bundle charts defined in the previous section. Let $s_\alpha:U_\alpha\rightarrow\pi^{-1}(U_\alpha)$ be a local section and $\phi_\alpha$ its associated bundle chart. Define the gauge field $A_\alpha:T(U_\alpha)\rightarrow\mathbb{R}$ and the Maurer-Cartan one-form $\theta_L:TS^1\rightarrow \mathbb{R}$ by
\begin{align}
A_\alpha&=s_\alpha^*\mathcal{A}\\
\theta_L(\theta,\xi)&=\xi,
\end{align}
where we have made the identification $TS^1=S^1\times\mathbb{R}$. Note that $\theta_L$ is nothing more than the coordinate differential on $S^1$. It is not difficult to show that on $\pi^{-1}(U_\alpha)$ $\mathcal{A}$ is made up of these two quantities according to
\begin{align}
\mathcal{A}=\pi^*A_\alpha+g_\alpha^*\theta_L.
\end{align}
This formula has two important consequences. First of all, if $A_\beta$ is another gauge field defined on an overlapping patch of $P/S^1$, $U_\alpha\cap U_\beta\neq\emptyset$, then it must be related to $A_\alpha$ on the overlap:
\begin{align}\label{gf}
A_\alpha= A_\beta+g_{\alpha\beta}^*\theta_L,
\end{align}
where $g_{\alpha\beta}:U_\alpha\cap U_\beta\rightarrow S^1$ is the circle-valued function defined by the relation $g_{\alpha\beta}(\pi(p))=g_\beta(p)-g_\alpha(p)$. Second, it implies that the \emph{gauge field strengths} $F_\alpha=dA_\alpha$, apparently only locally defined quantities, actually define a global two-form, the \emph{curvature form} $F$, over the entire base $P/S^1$. This result follows from applying the exterior derivative to \eqref{gf} and recalling that $d\theta_L=0$. On any of the $U_\alpha$, $F=F_\alpha$. As discussed in Ref. \onlinecite{chern}, the curvature two-form encodes the basic topological properties of the principal circle bundle it comes from.

Connection forms also provide a convenient structure for expressing the transformation law for the bundle chart representatives of globally defined vector fields on $P$. If $X:P\rightarrow TP$ is a smooth vector field on $P$, then given a bundle chart $\phi_\alpha$, its bundle chart representative is $X_\alpha\equiv\phi_{\alpha*}X:U_\alpha\times S^1\rightarrow TU_\alpha\times S^1\times\mathbb{R}$; the bundle chart representatives are just the vector field expressed in the coordinates provided by the bundle charts. Set $ X_\alpha(u,\theta)=(w_\alpha(u,\theta),\theta,\xi_\alpha(u,\theta))$, where $w_\alpha(u,\theta)\in T_u(P/S^1)$ and $\xi_\alpha(u,\theta)\in\mathbb{R}$. Using the fact that $\phi_\alpha^*X_\alpha=\phi_\beta^* X_\beta$ on $\pi^{-1}(U_\alpha\cap U_\beta)$, it is straightforward to show that the bundle chart representatives are related by
\begin{align}
w_\alpha(u,\theta)&=w_\beta(u,\theta^\prime)\\
\xi_\alpha(u,\theta)&=\xi_\beta(u,\theta^\prime)+g_{\beta\alpha}^*\theta_L(w_\beta(u,\theta^\prime)),
\end{align}
where $\theta^\prime=\theta+g_{\alpha\beta}(u)$. Using the transformation law for the gauge fields, this can be recast as
\begin{align}
\eta_\alpha(u,\theta)&\equiv\xi_\alpha(u,\theta)+A_\alpha(w_\alpha(u,\theta))\\
\label{c1}w_\alpha(u,\theta)&=w_\beta(u,\theta^\prime)\\
\label{c2}\eta_\alpha(u,\theta)&=\eta_\beta(u,\theta^\prime).
\end{align}
So we see that the $w_\alpha$ and $\eta_\alpha$ are local representatives of globally defined maps. To be precise, $w_\alpha=w\circ \phi_\alpha^{-1}$ and $\eta_\alpha=\eta\circ\phi_\alpha^{-1}$, where $w:P\rightarrow T(P/S^1)$ and $\eta:P\rightarrow\mathbb{R}$ are globally defined maps only constrained to satisfy $\tau_{P/S^1}\circ w=\pi$ ($\tau_{P/S^1}$ is the tangent bundle projection map associated to $T(P/S^1)$). 

Conversely, if there is an assignment of a local vector field $X_\alpha$ to each of the bundle charts $\phi_\alpha$ whose components satisfy \eqref{c1} and \eqref{c2}, then this collection of locally defined vector fields will define a global vector field $X:P\rightarrow TP$ that agrees with each of the $X_\alpha$ in the bundle charts. 

Why is expressing the vector transformation law in terms of the gauge fields useful? Because of the organization it brings to the process of stitching together local vector fields into a global one. The vector transformation law for passing from one arbitrary (non-bundle) coordinate chart to another would be quite messy to work with for this purpose. By working with the bundle charts and finding expressions for the gauge fields, the process is streamlined to finding the two functions $w$ and $\eta$. 

\section{Sufficiency of the flux condition}
The appropriate way to tackle this problem is to recognize that $SD$ is actually a principal circle bundle and that the existence of a globally defined perpendicular unit vector is equivalent to the existence of a global section of $SD$ (see appendix A for the necessary background on principal circle bundles). Because a principal circle bundle admits a global section if and only if it is a trivial bundle, the existence problem can be solved by appealing to the well-established topological classification of principal circle bundles \cite{chern}. This classification theorem tells us that if we can find any so-called principal connection on $SD$ (see appendix B for the necessary background on principal connections), which is a special sort of one-form over $SD$, then the curvature of this connection, a closed two-form over $D$ induced by the principal connection, will be exact if and only if $SD$ is a trivial bundle. Thus, given the curvature form, existence of global perpendicular unit vectors can be tested by integrating the curvature form over a collection of cycles that generate $D$'s second homology group $H_2(D,\mathbb{Z})$ \cite{bott}. If all of these integrals vanish, then the curvature form must be exact and a global section of $SD$ must exist. 

So in order to furnish a solution to the existence problem, all that we must still do is 1) prove that $SD$ is a principal circle bundle whose global sections, if they exist, coincide with global perpendicular unit vectors and  2) derive an expression for the curvature form associated to some principal connection on $SD$. Then existence can be determined in any particular case after finding the ``holes'' in $D$.

First notice that $SD$ is indeed a manifold. Actually it is a submanifold of $TD=D\times\mathbb{R}^3$ defined by the algebraic equations
\begin{align}
v\cdot v&=1\\
v\cdot b(x)&=0,\nonumber
\end{align}
where $(x,v)\in TD$. Next, consider the following circle action on $SD$:
\begin{align}
\Phi_\theta(x,v)=(x,\exp\left(\theta \hat{b}(x)\right)v),
\end{align}
where $\hat{b}(x)$ is the $3\times 3$ antisymmetric matrix defined by $\hat{b}(x)w=b(x)\times w$, and $\exp$ denotes the matrix exponential. Hence this circle action simply rotates all of the circles that comprise $SD$ by $\theta$ radians. Furthermore, the action is free. Therefore $(SD,\Phi)$ forms a principal circle bundle. 

To see that the sections of this circle bundle are equivalent to the perpendicular unit vectors, we first show that the base space of the bundle can be identified with $D$. Define the map $\pi:SD\rightarrow D$ by
\begin{align}
\pi(x,v)=x.
\end{align}
$\pi$ is a surjective submersion and $\pi^{-1}(x)$ is equal to the circle in $SD$ over $x$, which is an entire orbit of the action $\Phi$. It follows that $SD/S^1= D$ with $\pi$ serving as the bundle projection map. Thus a global section of $SD$ would consist of a smooth map of the form $s:D\rightarrow SD$ with the property $\pi(s(x))=x$, that is, $s(x)$ must lie in the circle over $x$. Because all of the points on the circle over $x$ are by definition perpendicular to $b(x)$ and of unit length, $s$ would be a global perpendicular unit vector. Conversely, any global perpendicular unit vector would define such an $s$. 

Now we move on to define a principal connection on $SD$. Because it will be necessary to work with the space $TSD\subseteq TTD$, we make the following identification:
\[
TTD=T(D\times\mathbb{R}^3)=TD\times T\mathbb{R}^3=(D\times\mathbb{R}^3)\times(\mathbb{R}^3\times\mathbb{R}^3).
\]
Accordingly, a typical element of the 12 dimensional space $TTD$ will be denoted $(x,u,v,a)$, where $(u,a)$ forms the tangent vector over the point $(x,v)\in TD$. Clearly, each element of $TSD$ can also be written in this way (of course $u$ and $a$ will be constrained in this case). It will also be helpful to define a metric on $TD$. Recall that such a metric on $TD$ defines an inner product on each of the tangent spaces in $TTD$. The useful metric in this case assigns an inner product to each $(x,v)\in TD$ given by
\begin{align}
\bigg<(x,u,v,a),(x,u',v,a')\bigg>=u\cdot u'+a\cdot a'.
\end{align}
Note the distinction between this inner product denoted by square brackets and the usual dot product between vectors in $\mathbb{R}^3$. Finally, a principal connection $\mathcal{A}:TSD\rightarrow\mathbb{R}$ can be defined by
\begin{align}
\mathcal{A}(x,u,v,a)=\bigg<(x,u,v,a),(x,0,v,b(x)\times v)\bigg>=a\cdot b(x)\times v.
\end{align}
The two defining properties of a principal connection (appendix B) are straightforward to check.

Next we derive an expression for the curvature form associated to $\mathcal{A}$. Because a local section $s_\alpha:U_{\alpha}\subseteq D\rightarrow \pi^{-1}(U_\alpha)$ must be of the form
\begin{align}
s_\alpha(x)=(x,e_1(x)),
\end{align}
where $e_1$ is a locally defined perpendicular unit vector, the gauge fields must be of the form
\begin{align}
A_\alpha(x,w)=s_\alpha^*\mathcal{A}(x,w)&=\bigg(w\cdot\nabla e_1(x)\bigg)\cdot b(x)\times e_1(x)\\
&\equiv w\cdot R(x).\nonumber
\end{align}
As the notation suggests, $R=\left(\nabla e_1\right)\cdot b\times e_1=\left(\nabla e_1\right)\cdot e_2$  is the well-known quantity that appears elsewhere in the guiding center formalism. Therefore, the curvature form $F=d A_\alpha$ is given by the equation
\begin{align}
\ast F=N\cdot dx,
\end{align}
where $\ast$ is the hodge star and $N=\nabla\times R$. By the transformation law for curvature forms given in appendix B, $N$ must be a globally defined quantity even when $e_1$, and therefore $R$, is not. In fact, there is an expression giving $N$ in terms of $b$ \cite{littlejohn81,littlejohn83}:
\begin{align}\label{N}
N=&\frac{1}{2}b\bigg(\text{Tr}(\nabla b\cdot \nabla b)-(\nabla\cdot b)^2\bigg)\\
&+(\nabla\cdot b)b\cdot\nabla b-b\cdot\nabla b\cdot \nabla b.\nonumber
\end{align}

With \eqref{N} in hand, all of the tools required to determine the existence of global perpendicular unit vectors have been assembled. To reiterate, to test for existence, the curvature form $F$ should be integrated over a collection of cycles that generate $D$'s second homology group $H_2(D,\mathbb{Z})$. Intuitively, this amounts to calculating the flux of $N$ through a collection of closed, bounded, boundary-less surfaces that encapsulate the ``holes'' in $D$. If all of these integrals vanish, then there will be global perpendicular unit vectors. Otherwise, owing to the ensuing non-trivial topology of $SD$, global perpendicular unit cannot be defined, even in principle.

\providecommand{\noopsort}[1]{}\providecommand{\singleletter}[1]{#1}%
%


\end{document}